\documentclass[prl,aps,floatfix,twocolumn]{revtex4}
\usepackage{latexsym,amssymb,graphicx}
\usepackage{epsfig}

\begin{document}

\title{Entropy and Temperature of a Static Granular Assembly}

\author{Silke Henkes$^{*}$, Corey S. O'Hern$^{\dagger}$ and Bulbul Chakraborty$^{*}$ }

\affiliation{
 $^{*}$ Martin Fisher School of Physics, Brandeis University, Waltham, MA 02454-9110 \\
 $^{\dagger}$ Department of Mechanical Engineering, Yale University,
New Haven, CT 06520-8284 and Department of Physics, Yale University, New Haven, CT 06520-8120
}

\begin{abstract}
Granular matter is comprised
of a large number of  particles  whose collective behavior determines macroscopic properties such as flow and mechanical strength\cite{kadanoffreview,NagelReview}.  A comprehensive theory of
the properties of granular matter, therefore,  requires a statistical framework. 
In molecular matter, equilibrium statistical mechanics\cite{chandlerbook}, which is founded on the principle of conservation of energy, provides this framework. 
Grains, however, are small but macroscopic objects whose interactions are dissipative since energy can be lost through excitations of the internal degrees of freedom.   
In this work, we construct a statistical framework for static, mechanically stable packings of grains,  which 
parallels that of equilibrium statistical mechanics but with conservation of energy replaced by the conservation of a function related to the 
mechanical stress tensor.  Our analysis demonstrates the existence of a state function that has all the attributes of entropy\cite{chandlerbook}.  In particular,  maximizing this 
state function leads to a well-defined granular temperature for these systems.
Predictions of the ensemble are verified against simulated packings of frictionless, deformable disks.
Our demonstration that a statistical ensemble can be constructed through
the identification of conserved quantities other than energy is a new approach that is expected to open up avenues for statistical descriptions of other non-equilibrium systems. 
\end{abstract}

\maketitle

Phases and phase transitions are manifestations of the collective behavior of many-particle systems. 
Phases are clearly delineated in atomic and molecular systems where solids resist shear but liquids do not.   In granular matter, however, phases lose some of their distinction\cite{kadanoffreview}:  
a pile of sand or coffee beans behaves as a solid but can break down in avalanches with only the surface layer flowing.  
Ordinary matter is a statistical ensemble of molecules with energy conserving interactions.  In such systems, equilibrium statistical mechanics
connects the microscopic properties to the observable, macroscopic behavior through probability distributions characterized by parameters such as temperature, and density\cite{chandlerbook}. 
Granular materials differ from their molecular counterparts in two important respects; 
(a) the grains are themselves macroscopic and lose energy every time they interact and (b) because of their size, gravitational effects swamp thermal fluctuations and changes are effected through mechanical acts such as 
shaking and tumbling.   The dissipative nature
of granular materials makes them explicitly non-equilibrium: energy has to be supplied to the system to maintain a steady state.  If the energy supply is turned off, granular materials relax to a static, mechanically
stable packing (a blocked state) through the dissipation of energy.  Response and phase behavior of weakly-driven granular systems are determined by the excitations of these blocked states. 
Prediction of these collective phenomena
entails making a statistical connection between the grain-level
and macroscopic properties of blocked states since, given a fixed set of macroscopic parameters,  e.g. volume and number of grains, many such states are possible .   

In this work, we construct a statistical framework for  predicting  probability
distributions of blocked states, which applies to frictional and frictionless grains but is limited to isotropic ones.   The detailed analysis is restricted to planar packings
where the mechanical stress tensor is known to have a simple structure\cite{BB,Timoshenko}.  The fundamental postulates, however, apply to higher dimensions.

The paper is organized as follows: (a) through the introduction of a height field for planar packings, we show that the total stress tensor is a conserved quantity; (b) we formulate a statistical theory
of blocked states which  parallels that of thermal systems but with the internal virial\cite{internalvirial-ref}, $\Gamma$, playing the role of energy; (c) we postulate entropy maximization to introduce 
a ``granular temperature'' 
and construct a probability distribution  involving the inverse granular temperature, $\alpha$ ; (d) we check this postulate against packings generated through a dynamical algorithm
by showing that all parts of a given packing have the same $\alpha$; (e) 
finally, we use the deduced form of $\alpha(\Gamma)$ to demonstrate that the entropy is indeed the logarithm of the density of states.  This last result  provides strong support for the microcanonical 
ansatz that all states with the same $\Gamma$ are equally likely and connects to the Edwards hypothesis for incompressible grains that all blocked states  with the same volume are equally likely\cite{Edwards-Mounfield}.

\begin{figure}
\scalebox{0.6}{\includegraphics{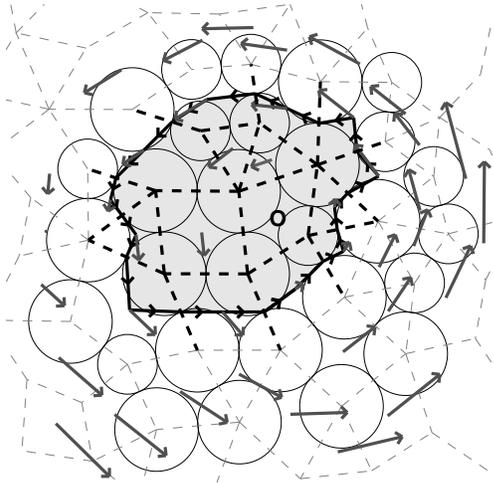}}
\caption{
\footnotesize{{\bf Height map:}
A height map with its origin at {\bf O} in the interior of a packing obtained from simulations.
The height vectors ${\vec h_{\mu}}$ (light arrows) are shown along with the boundary vectors ${\vec r}_{\mu 1}$, ${\vec r}_{\mu 2}$ (dark arrows)
entering eq. (\ref{boundary}). The dashed lines denote intergrain contacts and the ones 
included in  $\hat \sigma_{A}$ for a 8-grain cluster occupying the shaded region with area $A$, are in bold.}}

\label{Height}
\end{figure}

Every grain in a mechanically stable packing of frictional or frictionless grains
has to satisfy the constraints of force and torque balance.  
For planar packings of grains which interact only through contact forces, these constraints can be  incorporated  through a mapping to a set of auxiliary (height) variables
defined on the dual network of  the voids surrounded by grains\cite{BB,Chakraborty}.  Choosing the center of an
arbitrary void as the zero of the height field, the height vectors are constructed iteratively through: ${\vec h}_{\nu}={\vec F}_{ij}+{\vec h}_{\mu}$ 
where $ij$ is the contact traversed in going from the center of the void
$\mu$ to that of  $\nu$ ({\it cf}  Fig \ref{Height}).  Since the ${\vec F}_{ij}$'s  around a grain sum to zero,  the mapping of forces to heights is one to one up to an arbitrary choice of the origin ($O$ in Fig \ref{Height}).   
This mapping satisfies Newton's third law and enforces the force-balance constraint.  The constraint of torque balance imposes a divergenceless condition 
on the height field\cite{BB,Chakraborty}.   We use this mapping to demonstrate a conservation law related to the mechanical stress.

The height field, which  enforces that the stress tensor is divergence-free\cite{BB},  is an exact analog of the vector potential in electromagnetism, which ensures that 
the magnetic field is divergence-free\cite{Griffithsbook}.  As shown in the Methods section, 
the height map leads to an explicit implementation of Stokes' theorem\cite{Griffithsbook} which proves that the stress tensor $\hat \sigma_{A}$ 
can be expressed as  a sum involving only the boundary of the area $A$ occupied by the grains.   This result allows us to define an extensive quantity $\hat{\Sigma}= A \hat{\sigma}_{A}$ which
is left invariant by local rearrangements.   Different packings with the same value of $\hat{\Sigma}$ can be related to each other through processes that do not involve the boundary.
An example is provided by the ``wheel moves" in a triangular lattice\cite{Socolar}, which transposes one force-balanced packing into another leaving  $\hat{\Sigma}$ unchanged. 

In two dimensions,  the tensor $\hat{\Sigma}$ has two scalar invariants and these
can be taken to be the trace $\Gamma$, which is the internal virial\cite{internalvirial-ref}, and the determinant, $\kappa$.
For $ d \geq 3$, the loops around grains cannot be defined unambiguously and the height map cannot be constructed. 
The total stress tensor, $\hat{\Sigma}$, is nevertheless conserved under internal rearrangements since it can be written as the curl of a 2$^{nd}$ rank tensor\cite{EricKramer}, and Stokes' theorem can be used to 
represent it as a boundary integral. 

The above analysis demonstrates that for blocked states of frictional and frictionless packings of deformable disks,  $\Gamma = \sum_{ij} d_{ij} F_{ij}$ ({\it cf} Eq. \ref{micstress})  
is a pure boundary term. 
Since packings with different values of $\Gamma$ cannot be transformed into one another through internal rearrangements not involving the boundary, $\Gamma$ becomes the analog of energy in thermal systems.   It is then 
rigorously possible to calculate the density of states $\Omega(\Gamma, V,N)$  of  all blocked configurations
characterized by a given $\Gamma$, volume $V$ and number of grains $N$.  Analogous calculations, based on the conservation of physical variables other than energy,
have been performed in certain lattice models with rigid constraints\cite{Dasgupta}.
An entropy function can  be defined as $S = \ln \Omega(\Gamma, V,N)$ and assumption of 
an entropy maximization principle for granular equilibrium implies equalization of the granular temperature\cite{chandlerbook}:
$\alpha = \partial S /\partial \Gamma |_{V,N}$.  A constant-$\alpha$ canonical ensemble\cite{chandlerbook} follows with the probability of finding a blocked state $\nu$ being given by:
$P_{\nu} = \exp(-\alpha \Gamma_{\nu})/Z(\alpha, V, N)$.
In the current work, we ignore the invariant determinant $\kappa$.
It can easily be included as $S(\Gamma,\kappa,V,N)$ and gives rise to an additional factor $e^{-\mu \kappa}$ in the probability.   For frictionless systems, $\kappa$ is directly related to $\Gamma$.

\begin{figure}

\includegraphics[width=0.99\columnwidth]{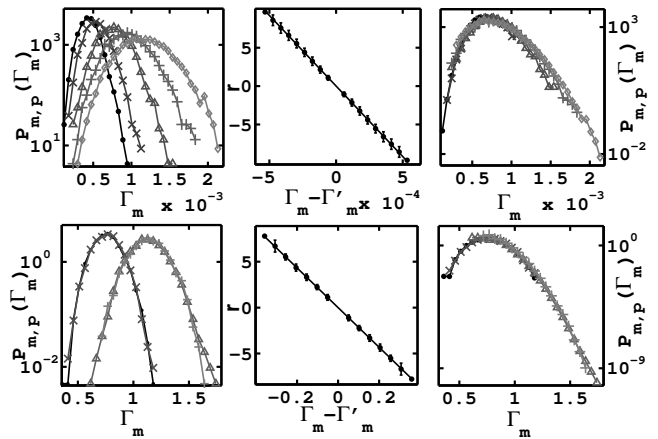}

\caption{\footnotesize{{\bf Numerical test of the equality of $\bf \alpha$:} Row 1: Results from packings with $N=4096$, $m=8$ and $\phi=0.838-0.844$. 
Row 2: Same for $\phi=0.95$ and $1.0$. First column shows $P_{m,p}(\Gamma_{m})$; second column is $r_{p,q}$ vs. $\Gamma_{m}-\Gamma'_{m}$ for an arbitrarily chosen pair $(p,q)$.
The error bars derive from the spread in $r_{p,q}$ obtained for different $\Gamma_{m}$ and $\Gamma'_{m}$, contributing to a given $\Gamma_{m}-\Gamma'_{m}$;
third column illustrates the scaling of $P_{m,p}(\Gamma_{m})$ according to Eq. \ref{rescalings}}}
\label{Panel}
\end{figure}


To test the predictions of our statistical framework, we
generate sets of mechanically stable packings of a fixed number $N$ of deformable
circular disks and group them according to their packing fraction $\phi = N v /V$,  where $v$ is the microscopic
grain volume and $V$ is the volume of the packing (see Methods section).
Each packing is characterized by $\Gamma_{N}$,  the total value of $\Gamma$ for the $N$-grain packing and by $\langle z \rangle$, the average
value of the number of contacts in the packing.  Units of length and interaction potential are chosen to make $\Gamma$ dimensionless (see Methods section).

The entropy maximization hypothesis implies\cite{chandlerbook} that all subregions of a grain packing ($p$), generated through
some dynamics, have the same granular temperature $\alpha_{p}$.   The values of
$\Gamma_{m}$ for $m$-grain clusters inside packing $p$ should then be distributed according to: 
$P_{m,p}(\Gamma_{m}) = \sum_{\nu}e^{-\alpha_{p} \Gamma_{\nu}} \delta(\Gamma_{\nu}-\Gamma_{m})$.  By measuring $P_{m,p}$ at a pair of values, $\Gamma_{m}$ and $\Gamma'_{m}$,
and comparing these to $P_{m,q}$ for the {\it same} pair we construct the ratio: 
\begin{equation} r_{p,q} \equiv \log \left( \frac{P_{m,p}(\Gamma_{m}) P_{m,q}(\Gamma'_{m})}{P_{m,p}(\Gamma'_{m}) P_{m,q}(\Gamma_{m})}\right) 
= -(\alpha_{p} - \alpha_{q})(\Gamma_{m} - \Gamma'_{m})  \label{logalpha1}
\end{equation}
The last equality  follows if  the packings $p$ and $q$ have the same or similar values of $\phi$ (see Methods section). 
Packings with $N=4096$ in two narrow packing fraction ranges are shown in the first column of Figure \ref{Panel}.  We find that for pairs $(p,q)$ of packings in the low ($0.838-0.844$) packing 
fraction range and in the range ($0.95, 1.00$), $r_{p,q}$ is indeed a linear function of $\Gamma_{m} - \Gamma'_{m}$,  establishing  that $\alpha$ is equal for all parts of a grain packing. 
An example of the linear correlation is shown in column 2 of 
Figure \ref{Panel}.
The average values of   $\alpha_{p}-\alpha_{q}=\delta \alpha _{p,q}$  extracted from these plots are $\approx 2 \times 10^{4}$ and $20$, 
for the low packing fractions and the high packing fractions, respectively.
A further test of the equality of $\alpha$ inside a packing is to verify that
for a pair $(p,q)$  with negligible differences in $\phi$, 
\begin{equation}
P_{m,q}(\Gamma_{m})\exp(-\delta\alpha_{p,q} \Gamma_{m}) = P_{m,p}(\Gamma_{m}) ~,
\label{rescalings}
\end{equation}
if both distributions are normalized to unity.
The third column in Figure \ref{Panel} demonstrates the scaling of packings with an arbitrary packing chosen as the reference $p$.

For packings with $\phi \sim 0.84$, it is possible to find overlapping distributions, $P_{m,p}(\Gamma_{m})$  over a range of $\Gamma_{N}$ extending from $\simeq 10^{-5} - 10^{-3}$.
In this regime, we extract $\delta \alpha _{p,q}$ systematically from Eq. \ref{rescalings} by rescaling the same packing $q$ with two different reference packings, $p$,  
and determine  $\alpha$  as a function of $\Gamma_{N}$, up to an arbitrary constant $C$. For $m>2$,  we find that over two orders of magnitude in $\Gamma_{N}$, the results are independent of $m$ and
described by the functional form:
\begin{equation}
\alpha=C+2 N/\Gamma_{N}
\label{alphagamma}
\end{equation}
Eq. \ref{alphagamma} implies that the granular temperature $1/\alpha$ tends to 0 for $\Gamma_{N}\rightarrow 0$.  The value of $C$ is determined by the
behavior of $\alpha$ for large $\Gamma_{N}$ .  Since we expect $1/\alpha \rightarrow \infty$ as $\Gamma_{N} \rightarrow \infty$, we choose $C=0$.

The above numerical tests demonstrate that for all the packings studied,  over the full range of $\phi$, different subregions of a packing have the same value of $\alpha$. 
Moreover, for $\Gamma_{N} \leq 10^{-3}$,  $\alpha$  is independent of $m$ and a unique function of $\Gamma_{N}$.  These results 
establish $\alpha$ as the granular equivalent of inverse temperature.  The parallel with equilibrium statistical mechanics is further illustrated below through
the verification of the analog of the Boltzman entropy formula\cite{chandlerbook} $S \propto \ln \Omega$ providing support for the microcanonical ansatz of 
equal probability of blocked states with the same $\Gamma$.

\begin{figure}
\begin{center}
\scalebox{0.45}{\includegraphics{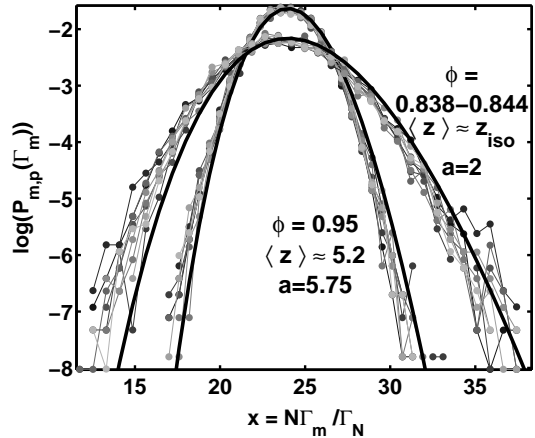}}
\caption{\footnotesize{{\bf Test of Eq. \ref{DOS}:} Distributions for $N=4096$ and $m=24$ plotted as a function of $x=N\Gamma_{m}/\Gamma_{N}$. 
For $\phi=0.838$ to $\phi=0.844$, distributions are compared to Eq. \ref{DOS}  with $a=2$ (solid line). For $\phi=0.95$, with $\langle z \rangle$ in the range $5.207\pm0.06$, Eq. \ref{DOS} 
with $a=5.75$ (solid line) best describes the data. }}
\label{scalings2}
\end{center}
\end{figure}

Since $\alpha$ is defined by the derivative of the entropy with respect to $\Gamma$,  Eq. \ref{alphagamma} implies a specific form of the density of states: 
$\Omega (\Gamma,V,N) = e^{S} \sim \Gamma^{Na}$,  with $a=2$, independent of $\phi$, for packings with $\phi \simeq 0.84$.  Combining this relation with the
functional form of $\alpha(\Gamma)$, it 
is straightforward to show that the distribution of $\Gamma_{m}$ in a packing ({\it cf} Eq. \ref{disbn})
is a function only of $x=N \Gamma_{m} /\Gamma_{N}$:
\begin{equation} \label{DOS} P_{m,p}(\Gamma_{m}) \equiv P_{m} (x) = C x^{m a} exp(-ax) ~,\end{equation}
The scaling of the distributions, obtained from multiple packings with $\phi \sim 0.84$, is illustrated in Figure \ref{scalings2} 
for $m=24$ and verifies that $S$ is indeed proportional to $\ln \Omega$ in this regime.

\begin{figure}
\begin{center}
\scalebox{0.45}{\includegraphics{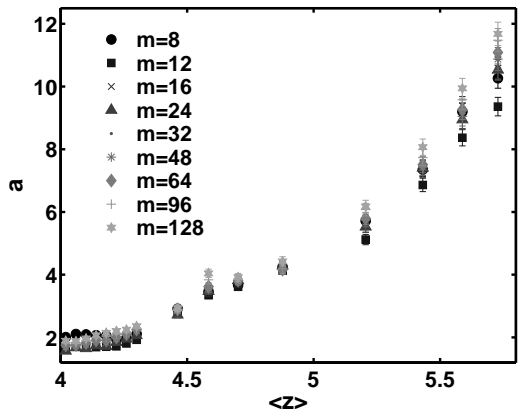}}
\caption{\footnotesize{{\bf Numerical test of the model for $\bf a$:}
Fitting parameter $a$ obtained from distributions of $N=1024$ grain-packings
with different values of the average number of contacts, $\langle z \rangle$.}}
\label{scalings3}
\end{center}
\end{figure}

For packings with $\phi \geq 0.85$,  we found that the distributions obtained from different packings no longer scale with $x=N \Gamma_{m}/\Gamma_{N}$
but depend explicitly on $\phi$; becoming narrower with increasing $\phi$.  
We know from the simulation data and previous work\cite{Nagel} that the average number of contacts, $\langle z \rangle$, is a strongly increasing function of the packing fraction $\phi$ and that, 
in the low packing fraction regime, 
$ \langle z \rangle \rightarrow z_{iso}=4$ (for frictionless disks
the isostatic packing has $2d$ contacts where $d$ is the spatial dimension\cite{Mouzarkel,Witten}).  It is also clear from Eq. \ref {DOS} that $1/a$ is proportional to the width of the rescaled distributions.
Based on these observations we hypothesize that the distributions are still described by Eq. \ref{DOS} but with $a=2+f(\langle z \rangle-z_{iso})$ and  $f(0)=0$. A test of this hypothesis is presented in
Figure \ref{scalings2}.   For packings with $\phi=0.95$,  the measured $\langle z \rangle\approx 5.2$ and the best fit to Eq. \ref{DOS} yields $a=5.75$.
For smaller systems, with $N=1024$,  we performed numerical fits to the distributions
over a wide range of $\langle z \rangle$.
Figure \ref{scalings3} shows the numerical values of $a$ for a discrete set of $\langle z \rangle$ values and for $8 \leq m\leq 128$. 
It is clear that $a$ increases with $\langle z \rangle$ and approaches a constant for $\langle z \rangle \rightarrow 4$.

The results presented above show that the granular temperature $1/\alpha$ of a $N$-grain packing is determined by $\langle z \rangle$ and $\Gamma_{N}$:  $\alpha \simeq N a(\langle z \rangle)/\Gamma_{N}$.   
This relationship is the analog of an equation of state relating energy, temperature and density in a thermal system.
Since $a$ approaches a constant as $\Gamma_{N} \rightarrow 0$  and $\langle z \rangle \rightarrow z_{iso}$, in this limit, the density of states is, to a good approximation,
only a function of $\Gamma$: 
\begin{equation}
\Omega(\Gamma,V,N)  \sim \Gamma^{2N} \propto  F^{2N}~.
\label{decoupled}
\end{equation}
Here $F = \sum_{ij} F_{ij} $ is the sum of the magnitude of all the contact forces and 
we have used the fact that as $\Gamma \rightarrow 0$, the compression of the grains approach zero and, in this limit,  $\Gamma =\sum_{ij} d_{ij} F_{ij}\propto F$.
The statistical ensemble of packings in this incompressible limit
should be identical to that of infinitely rigid grains since, as we have demonstrated, the ensemble is determined by $\Gamma_{N}$. 
The conservation of $\Gamma$ reduces to a conservation of $F$ for infinitely rigid grains and we recover the Edwards-microcanonical
ensemble based on total volume and total external force\cite{Edwards-Mounfield,vanHecke,Edwards-private}.  Such an ensemble would lead to a density of states of the form (\ref{decoupled}).

In summary, we have shown that it is possible to construct a statistical ensemble for mechanically stable (blocked) states of frictional and frictionless granular packings.  
Numerical tests demonstrate the existence
of a granular temperature which is given by the logarithmic derivative of the density of states with respect to $\Gamma$.  The tests also establish that the density of states is history-independent 
since different 
paths leading to the same $\Gamma_{N}$ generate the same distributions.  Our analysis was performed on data
obtained from a specific though fairly generic dynamics with no {\it a priori} information about the ensemble and, therefore we expect our results to be applicable in general.

The statistical framework presented in this paper leads to a natural
phase space for jamming\cite{Liu} with the axes being thermodynamic temperature ($T$),  packing fraction ($\phi$), and granular temperature $1/\alpha$. 
Simulations\cite{Nagel} and experiments\cite{Trush} have indicated the existence of a $T=0$ critical point (Point J) on the $\phi$ axis.  
The $\alpha$ ensemble  
provides a framework for exploring phases, phase transitions and critical points on the $T=0$ plane. 
A preliminary attempt, based on the present framework but without any rigorous justification of its
fundamental premise, has been made by two of us\cite{Chakraborty} and has led to a field theory of Point J.   One of the crucial aspects of this field theory is the coupling 
between $\langle z \rangle$ and $\Gamma$.
Our current analysis shows that this coupling is a consequence of the nature of the density of states and that the geometry and stress decouple only in the infintely-rigid-grain or the 
incompressible limit ({\it cf} Eq. \ref{decoupled}); {\it i.e.} at Point J. 

The full richness of granular statistical mechanics, which is founded on a conserved tensor not a scalar, can be explored only in systems with friction.
Our tests of the statistical ensemble so far have been confined to numerical simulations of frictionless disks because the verification required extensive statistical sampling.  
We are in the process of extending the verification of the statistical ensemble
to experimental packings of grains with friction\cite{Behringer-private}.\\

\noindent{\bf Methods}\\

\footnotesize{\noindent{\bf Height Map}
The microscopic stress tensor for grain $i$ is defined as\cite{Landau} 
\begin{equation} \hat{\sigma}_{i} = \sum_{j=1}^{z_{i}} \vec{d}_{ij} \vec{F}_{ij}, \label{micstress} \end{equation}
where the $\vec{d}_{ij}$ are the vectors connecting the grain center to its contact points, the $\vec{F}_{ij}$ are the contact forces and $z_{i}$ is the number of contacts for grain $i$.  The height map
can be used to rewrite $\hat{\sigma}_{i}$ as\cite{BB}:
$\hat{\sigma}_{i} = \sum_{\mu =1}^{z_{i}} (\vec{r}_{\mu 1} + \vec{r}_{\mu 2} ) \vec{h}_{\mu}$, where the $\vec{r}_{\mu 1}$ and $\vec{r}_{\mu 2}$ are the vectors linking the two
contact points associated with the the void $\mu$ ({\it cf} Fig \ref{Height}). The grain area is defined as the area enclosed by the $\vec{r_{\mu}}$-vectors, and these tesselate the plane ({\it cf} Fig \ref{Height}).
The macroscopic stress tensor $\hat{\sigma}$ 
is related to its single grain counterpart through $\hat{\sigma}_{A} = \frac{1}{A} \sum_{i\in A} \hat{\sigma}_{i}$, where the summation is over a connected cluster of grains and the area $A$ is the 
combined grain area of all grains in the cluster.
In the summation over grains in $\hat{\sigma}_{A}$,
the terms related to voids fully inside the area $A$ add up to zero and, as illustrated in Fig \ref{Height}, we are 
left with a sum over the voids at the boundary of $A$ : 
\begin{equation}
\hat{\sigma}_{A} = \frac{1}{A} \sum_{\mu \in boundary}(\vec{r}_{\mu 1} + \vec{r}_{\mu 2} ) \vec{h}_{\mu} ~ .
\label{boundary}
\end{equation} \\
\noindent{\bf Simulations} 
We generate mechanically stable
packings of bidisperse deformable disks that interact via purely repulsive linear spring
interactions 
for systems with $N=1024$ and $N=4096$ disks.  The mixtures are 50-50
by number and the diameter ratio between large and small disks
is 1.4.
For each of these sizes we study several packing fractions from 
around random close packing $\phi=0.84$ to more than $20\%$ above this value.
To create the packings, we initialize the disks with random initial
conditions at a specified packing fraction and then implement
conjugate gradient energy minimization to find the nearest local
energy minimum\cite{Nagel}.  Near random close packing, the packings typically have some grains
with no contacts (``rattlers'').   In all of our analysis, these rattlers have been excluded.
 In the simulations, lengths are measured in units of the large-particle diameter and energy is measured in units of the characteristic interaction strength\cite{Nagel}.  This renders $\Gamma$ dimensionless and the contact force is numerically equal to the magnitude of the disk overlap.\\
\\
\noindent{\bf Testing equality of granular temperature} 
We divide a packing $p$ into subregions containing $m$ grains 
and construct the probability distribution $P_{m}(\Gamma_{m})$ by histogramming $\Gamma_{m} = \sum_{ij \subseteq m} r_{ij} F_{ij}$ is summed over all contacts
in the subregion (including boundary contacts).\\
The prediction of the  canonical $\alpha$-ensemble is that this distribution should have the form 
\begin{eqnarray}
P_{m,p} (\Gamma_{m})& \equiv &\sum_{\nu}e^{-\alpha_{p} \Gamma_{\nu}} \delta(\Gamma_{\nu}-\Gamma_{m}) \nonumber \\
&= &\Omega (\Gamma_{m},v_{m},m) \exp (-\alpha_{p} \Gamma_{m})/Z_{m}(\alpha_{p})
\label{disbn}
\end{eqnarray}
where $\Omega (\Gamma_{m}, v_{m},m)$ is the density of states corresponding to  $v_{m}$, the volume of the subregion and $Z_{m}(\alpha_{p})$ is the partition function\cite{chandlerbook}.  
In practice, we evaluate $P_{m,p} (\Gamma_{m})$ by combining all $m$-grain regions irrespective of
the value of $v_{m}$.  The appropriate distribution to compare the numerical data to is, therefore,
\begin{equation} 
P_{m,p} (\Gamma_{m}) = \Omega^{eff}_{m}(\Gamma_{m}, V_{p},N) \exp (-\alpha_{p} \Gamma_{m})/Z_{m}(\alpha_{p})
\label{Pmphi} 
\end{equation} 
where $\Omega_{eff}$ depends on the volume $V_{p}$ of the packing $p$ but not on the local geometry.
For a given pair of packings with with the same $N$ and packing fraction differences small enough such that the $V$ dependence of   $\Omega^{eff}$ can be ignored, 
\begin{equation} 
r_{p,q} \equiv \log \left( \frac{P_{m,p}(\Gamma_{m}) P_{m,q}(\Gamma'_{m})}{P_{m,p}(\Gamma'_{m}) P_{m,q}(\Gamma_{m})}\right) 
\simeq  -(\alpha_{p} - \alpha_{q})(\Gamma_{m} - \Gamma'_{m})  
\label{logalpha}
\end{equation} 
since the factors of $\Omega_{eff}$ from the numerator and denominator cancel out.}\\
\\

\noindent{\bf Acknowledgments} 
BC wishes to thank Daan Frenkel for suggesting the
method for extracting $\alpha$ from numerical simulations. The authors would also like to
thank Bob Behringer and Mike Cates  for many useful discussions.   This work was supported by the National Science Foundation.

\bibliography{references-nature}

\end{document}